\documentclass[12pt, english]{article}

\usepackage[english]{babel}  % solve the error: You haven't loaded the option english yet
\usepackage[letterpaper, margin=1in]{geometry}  % margin, paper size etc.
\usepackage{bm}\usepackage{bbm}   % bm for bold math, bbm for \mathbbm{1}
\usepackage{amsmath}   % for \overset, \underset, \text, \lVert, \rVert
\usepackage{amssymb}   % amssymb or amsfonts for \mathbb
\usepackage{graphicx}  % \includegraphics, \resizebox. graphicx is better than graphics.
\usepackage{cases}     % \begin{cases} for stack of equations after the bracket
\usepackage{multirow}  % \multirow and \multicolumn
\usepackage{longtable} % \begin{longtable} for long tables.
\usepackage{caption}   % \caption*{}, \captionof (while using \resize)
\usepackage{float}     % prevent table from floating by \begin{table}[H]. Using [h] without this package does not avoid the floating.
\usepackage{adjustbox} % \begin{adjustbox} to resize the table
\usepackage[title]{appendix} % more control of appendix
\usepackage{array}     % vertical center in the table \begin{tabular}{ | m{1in} |... }
\usepackage{xcolor}    % change text color
\usepackage{setspace}   %Allows double spacing with the \doublespacing command
\usepackage{url}       % include url: \url{}

\newcommand{\vone}{\mathbbm{1}}      % command to write vector 1 in vector form
\newcommand{\norm}[1]{\left\| #1 \right\|}   % command to write norm
\usepackage{enumitem}  % Change label: \begin{enumerate}[label=(\arabic*)] . Not compatible with lyx: Remove [label=(\arabic*)].
\usepackage{hhline}    % \hhline double lines in tables. Not compatible with lyx: Replace \\hhline\{.*\}\} with \\hline, replace \\hhline\{.*\} with \\hline.

\begin{document}
\title{\vspace{-2cm}
Time-Invariance Coefficients Tests with the \\
Adaptive Multi-Factor Model\thanks{We thank Dr. Manny Dong, and all the supports
from Cornell University. Declarations of interest: none.}}
\author{ Liao Zhu\thanks{Ph.D. in the Department of Statistics and Data Science, Cornell University,
Ithaca, New York 14853, USA. Email: lz384@cornell.edu.} ,
Robert A. Jarrow\thanks{Corresponding Author. Ronald P. and Susan E. Lynch Professor of Investment Management, Samuel Curtis Johnson Graduate School of Management, Cornell University, Ithaca, New York 14853,
USA and Kamakura Corporation, Honolulu, Hawaii p96815. Email: raj15@cornell.edu.} ,
Martin T. Wells\thanks{Charles A. Alexander Professor of Statistical Sciences, Department
of Statistics and Data Science, Cornell University, Ithaca, New York 14853, USA. Email: mtw1@cornell.edu.}}
\date{}

\maketitle

\global\long\def\thefootnote{\arabic{footnote}}%
 %\blindtext

%\tableofcontents

\begin{abstract}
The purpose of this paper is to test the time-invariance of the beta coefficients estimated by the Adaptive Multi-Factor (AMF) model. The AMF model is implied by the generalized arbitrage pricing theory (GAPT), which implies constant beta coefficients. The AMF model utilizes a Groupwise Interpretable Basis Selection (GIBS) algorithm to identify the relevant factors from among all traded ETFs. We compare the AMF model with the Fama-French 5-factor (FF5) model. We show that for nearly all time periods with length less than 6 years, the beta coefficients are time-invariant for the AMF model, but not for the FF5 model. This implies that the AMF model with a rolling window (such as 5 years) is more consistent with realized asset returns than is the FF5 model.
\end{abstract}

%\tableofcontents

\clearpage{}

\section{Introduction}

The purpose of this paper is to test the a more general multi-factor beta model implied by the generalized arbitrage pricing theory (GAPT)\footnote{It is shown in Jarrow and Protter (2016) \cite{jarrow2016positive}
that both Merton's (1973) \cite{merton1973intertemporal} intertermporal
capital asset pricing model and Ross's APT (1976) \cite{ross1976arbitrage}
is a special case of the GAPT.} of Jarrow \& Protter (2016) \cite{jarrow2016positive}, which implies
constant beta coefficients. Estimating multi-factor models with time
varying betas is a difficult task. Consequently, for tractability,
the \emph{additional assumption} of time invariant betas in multi-factor
models is often employed, see Jagannathan et al. (2010) \cite{jagannathan2010cross}
and Harvey et al. (2016) \cite{harvey2016and} for reviews. The assumption
of constant betas may not be restrictive if the time horizon is short,
but for many applications restricting to short time horizons is problematic.
Alternative approaches for fitting time varying betas have been proposed
and estimated, such as the conditional factor models (see Adrian et
al. (2015) \cite{adrian2015regression}, Cooper \& Maio (2019) \cite{cooper2019new},
and Avramov \& Chordia (2006) \cite{avramov2006asset}).

In contrast to these approaches, we show herein that this constant
beta assumption can be avoided by using the GAPT. To estimate the
GAPT, we employ the newly-developed Adaptive Multi-Factor (AMF) model with the Groupwise
Interpretable Basis Selection (GIBS) algorithm to choose the relevant
factors from among all traded ETFs. The AMF model is the name given
to the methodology employed to estimate the beta coefficients using
LASSO regression from a population of potential factors identified
with the GIBS algorithm.

Using the AMF model with the GIBS algorithm proposed in Zhu et al.
(2020) \cite{zhu2020high}, but fitting price differences instead
of returns, we estimate a multi-factor model with constant beta coefficients
to equities over the time period 2007 - 2018. Employing the collection
of Exchange Traded Funds (ETFs) as potential factors, we use a high
dimensional GIBS algorithm to select the factors for each company.
No-arbitrage tests confirm the validity of the GAPT. As a robustness
check, we also show that the estimated model performs better than
the traditional Fama-French 5-factor (FF5) model. After this validation,
we perform the time-invariance tests for the $\beta$ coefficients
for various time periods. We find that for time periods of no more
than 5 years, the beta coefficients are time-invariant for the AMF
model, but not for the FF5 model. These results confirm
that using a dynamic AMF model with a rolling window length no more
than 5 years provides a better fit to equity prices compared to the FF5 model.

An outline for this paper is as follows. Section \ref{sec: ti_generalized_apt}
presents the GAPT theory. Section \ref{sec: ti_estimation} provides
the estimation methodology. Section \ref{sec: ti_test_methodology}
provides the testing methodologies and results. Section \ref{sec: ti_conclude}
concludes.

\section{The GAPT}

\label{sec: ti_generalized_apt}

Jarrow \& Protter (2016) derive a testable multi-factor model over
a finite horizon {[}0, T{]} in the context of a continuous time, continuous
trading market assuming only frictionless and competitive markets
that satisfy no-arbitrage and no dominance, i. e. the existence of
an equivalent martingale measure. As in the traditional asset pricing
models, adding a non-zero alpha to this relation (Jensen's alpha)
implies a violation of the no-arbitrage condition.

The GAPT uses a linear algebra framework to prove the existence of
an algebraic basis in the security's payoff space at some future time
$T$. Since this is is a continuous time and trading economy, this
payoff space is infinite dimensional. The algebraic basis at time
$T$ constitutes the collection of tradeable basis assets and it provides
the multi-factor model for a security's price at time $T$.\footnote{An algebraic basis means that any risky asset's return can be written
as a linear combination of a \emph{finite} number of the basis asset
returns, and different risky assets may have a \emph{different finite}
combination of basis asset explaining their returns. Since the space
of random variables generated by the admissible trading strategies
is infinite dimensional, this algebraic basis representation of the
relevant risks is parsimonious and sparse.} The coefficients of the time T multi-factor model are constants (non-random).
No-arbitrage, the existence of the martingale measure, implies that
the arbitrage-free prices of the risky assets at all earlier dates
$t\in[0,T)$ will satisfy the same factor model and with the same
constant coefficients. Transforming prices into returns (dividing
prices at time $t+1$ by time $t$ prices to get the return at $t+1$),
makes the resulting coefficients in the multi-factor model stochastic
when viewed at time $0$. However, this is not the case for the multi-factor
model specified in a security's price (or price differences). The
multi-factor model's beta coefficients in the security's price process
are time-invariant.

The GAPT is important for industry practice because it provides an
exact identification of the relevant set of basis assets characterizing
a security's \emph{realized} (emphasis added) returns. This enables
a more accurate risk-return decomposition facilitating its use in
trading (identifying mispriced assets) and for risk management. Taking
expectations of this realized return relation with respect to the
martingale measure determines which basis assets are \emph{risk-factors},
i.e. which basis assets have non-zero expected excess returns (risk
premiums) and represent systematic risk. Since the traditional models
are nested within the GAPT, an empirical test of the GAPT provides
an alternative method for testing the traditional models as well.

Let $B(t)$ denote the time $t$ value of a money market account (mma)
with initial value of \$1 at time 0, i.e.
\begin{equation}
B(t)=1\cdot\prod_{k=0}^{t-1}\Big[1+r_{0}(k)\Big]\label{eq: ti_mma}
\end{equation}
where $r_{0}(k)$ is the default-free spot rate (the risk-free rate)
from time $k$ to time $k+1$.

Let $A_{i}(t)$ denote the market price of the $i^{th}$ stock at
time $t$ for $i=1,\ldots,N_{c}$. To include cumulative cash flows
and stock splits into the valuation methodology, we need to compute
the adjusted price\footnote{This is sometimes called the gains process from investing in the security
over $[0,T]$.} $Y_{i}$, which is reconstructed by the using the security's returns,
after being adjusted for dividends and stock splits\footnote{These are the returns provided in the available data bases. }:
\begin{equation}
Y_{i}(t)=A_{i}(0)\prod_{k=0}^{t-1}\Big[1+R_{i}(k)\Big]\label{eq: ti_prc}
\end{equation}
where $A_{i}(0)$ is the initial price and $R_{i}(k)$ its return
over $[k,k+1]$.

Let $V_{j}(t)$ be the adjusted price of the $j^{th}$ basis asset
at time $t$ for $j=1,\ldots,N_{f}$. Here, the sources of the basis
asset in our model are the Fama-French 5 factors and the Exchange-Traded
Funds (ETFs). We include within this set of risk-factors the mma. For
notational simplicity, we let $V_{1}(t)=B(t)$.

Given this notation, the generalized APT implies the following multi-factor
model for the $i^{th}$ security's time $t$ price:
\begin{equation}
Y_{i}(t)=\sum_{j=1}^{N_{f}}\beta_{i,j}V_{j}(t)+\epsilon_{i}(t)\label{eq: ti_theory}
\end{equation}
where $\beta_{i,j}$ for all $i,j$ are constants and $\tilde{\epsilon}_{i}(t)$
is an i.i.d. error term with zero mean and constant variance. The key implication here is that the beta coefficients as implied by the GAPT are constants. This is an implication of the GAPT and not an additional
assumption.

The goal of our estimation is two-fold. First, we want to test to
see whether expression (\ref{eq: ti_theory}) provides a good fit
to historical stock price data. Second, we want to investigate whether
the multi-factors coefficients $\beta_{i,j}$'s are time-invariant, i.e. consistent with the GAPT. Given that prices are known to be autocorrelated, instead of estimating
expression (\ref{eq: ti_theory}), we fit the first order price differences
of this expression:
\begin{equation}
\Delta Y_{i}(t)=Y_{i}(t+1)-Y_{i}(t)=\sum_{j=1}^{N_{f}}\beta_{i,j}\Delta V_{j}(t)+\Delta\epsilon_{i}(t)\label{eq: ti_theory_diff}
\end{equation}
where $\Delta V_{j}(t)=V_{j}(t+1)-V_{j}(t)$ and $\Delta\epsilon_{i}(t)=\epsilon_{i}(t+1)-\epsilon_{i}(t)$.

For a given time period $(t,T)$, letting $n=T-t+1$, $p=N_{f}$ we
can rewrite the expression (\ref{eq: ti_theory}) using time series
vectors as
\begin{equation}
\bm{Y}_{i}=\bm{V}\bm{\beta}_{i}+\bm{\epsilon}_{i}\label{eq: ti_theory_vec}
\end{equation}
where $\bm{V}=(\bm{V}_{1},\bm{V}_{2},\ldots,\bm{V}_{p})$ and
\[
\bm{Y}_{i}=\begin{pmatrix}Y_{i}(t)\\
Y_{i}(t+1)\\
\vdots\\
Y_{i}(T)
\end{pmatrix}_{n\times1},\;\bm{V}_{j}=\begin{pmatrix}V_{j,t}\\
V_{j,t+1}\\
\vdots\\
V_{j,T}
\end{pmatrix}_{n\times1},\bm{\beta}_{i}=\begin{pmatrix}\beta_{i,1}\\
\beta_{i,2}\\
\vdots\\
\beta_{i,p}
\end{pmatrix}_{p\times1},\;\bm{\epsilon}_{i}=\begin{pmatrix}\epsilon_{i}(t)\\
\epsilon_{i}(t+1)\\
\vdots\\
\epsilon_{i}(T)
\end{pmatrix}_{n\times1}.
\]
Taking the first-order difference of each vector, the equation (\ref{eq: ti_theory_diff})
can be rewritten as
\begin{equation}
\Delta\bm{Y}_{i}=\Delta\bm{V}\bm{\beta}_{i}+\Delta\bm{\epsilon}_{i}\label{eq: ti_vec_diff}
\end{equation}
where $\Delta\bm{V}=(\Delta\bm{V}_{1},\Delta\bm{V}_{2},\ldots,\Delta\bm{V}_{p})$
and
\[
\Delta\bm{Y}_{i}=\begin{pmatrix}\Delta Y_{i}(t)\\
\Delta Y_{i}(t+1)\\
\vdots\\
\Delta Y_{i}(T-1)
\end{pmatrix}_{(n-1)\times1},\;\Delta\bm{V}_{j}=\begin{pmatrix}\Delta V_{j,t}\\
\Delta V_{j,t+1}\\
\vdots\\
\Delta V_{j,T-1}
\end{pmatrix}_{(n-1)\times1}
\]

\[
\bm{\beta}_{i}=\begin{pmatrix}\beta_{i,1}\\
\beta_{i,2}\\
\vdots\\
\beta_{i,p}
\end{pmatrix}_{p\times1},\;\Delta\bm{\epsilon}_{i}=\begin{pmatrix}\Delta\epsilon_{i}(t)\\
\Delta\epsilon_{i}(t+1)\\
\vdots\\
\Delta\epsilon_{i}(T-1)
\end{pmatrix}_{(n-1)\times1}.
\]

Based on the generalized APT theory, we will estimate the unknowns and test on
the time-invariance of the $\beta$'s using the AMF model with the GIBS algorithm (Zhu et al. \cite{zhu2020high}) in the following sections. The AMF model is the name given to the methodology
employed to estimate the beta coefficients using LASSO regression from a population of potential factors identified with the GIBS algorithm.

\section{The Estimation Methodology}

\label{sec: ti_estimation}

This section gives the estimation and the testing methodology. We
first specify the data used to estimate and to test the model. Then,
we pick time periods with various lengths with a sufficient quantity
of ETFs available. For each time period, we
use the GIBS algorithm to estimate the AMF model. Within each time period,
we propose several methods to test if the $\beta$'s are constant.
We will also compare the results for AMF with the benchmark FF5 model.

\subsection{Data and Time Periods}

\label{sec: ti_data}

We use the ETFs to select the basis assets
to be included in the multi-factor model. For comparison to the literature
and as a robustness test, we include the Fama-French 5 factors into
this set\footnote{For the FF5 factors, we add back the risk-free rate
if it is already subtracted in the raw data as in the market return
in Fama-French's website \url{https://mba.tuck.dartmouth.edu/pages/faculty/ken.french/data_library.html}.}.

The data used in this paper consists of all the stocks and all the
ETFs available in the CRSP (The Center for Research in Security Prices)
database over the years 2007 - 2018. We start in 2007 because prior
the number of ETFs with a sufficient amount of capital (in other word,
tradable) is limited, which makes it difficult to fit the AMF model.

To avoid the influence of market microstructure effects, we use a
weekly observation interval. A security is included in our sample
only if it has prices and returns available for more than 2/3 of all
the trading weeks. This is consistent with the empirical asset pricing
literature. Our final sample consists of over 4000 companies listed
on the NYSE. To avoid a survivorship bias, we include the delisted
returns (see Shumway (1997) \cite{shumway1997delisting} for more
explanation). For each regression time period, we form the adjusted
price using Equation (\ref{eq: ti_prc}).

We repeat our analysis on all sub-periods of 2007 - 2018 (both ends
inclusive, the same below) starting at Jan. 1st, ending at Dec. 31th,
and with a length at least 3 years. This means we have 55 sub-periods
in total, such as 2007 - 2009, 2008 - 2010, ..., 2008 - 2018, 2007
- 2018. We only use the time periods with length at least 3 years
to ensure that there are sufficient observations to fit the models.
As stated above, we use the weekly observations to avoid the market
microstructure effects, so 3 years means $n\geq156$ observations
in such regression.

The number of ETFs in our sample is large, slightly over 2000 by the
end of 2018, i.e. the number of basis assets $p_{1}>2000$. The set
of potential factors is denoted $V_{j}(1\leq j\leq p_{1})$. Since
$n=156<2000<p_{1}$ for the 3-year time periods, high-dimensional
statistical methods, including the GIBS algorithm, need to be used.

\subsection{High-Dimensional Statistics and the GIBS Algorithm}

This section provides a brief review of the high-dimensional statistical
methodologies used in this paper, including the GIBS algorithm.

Let $\norm{\bm{v}}_{d}$ denote the standard $l_{d}$ norm of a vector,
then
\begin{equation}
\norm{\bm{v}}_{d}=\left(\sum_{i}|\bm{v}_{i}|^{d}\right)^{1/d}\qquad\text{where}\;\;0<d<\infty,
\label{eq: norm}
\end{equation}
specifically, $\norm{\bm{v}}_{1}=\sum_{i}|\bm{v}_{i}|$. Suppose $\bm{\beta}$
is a vector with dimension $p\times1$. Given a set $S\subseteq\{1,2,...,p\}$,
we let $\beta_{S}$ denote the $p\times1$ vector with i-th element
\begin{equation}
(\bm{\beta}_{S})_{i}=\begin{cases}
\beta_{i}, & \text{if}\;\;i\in S\\
0, & otherwise.
\end{cases}\label{eq: set_index}
\end{equation}
Here the index set $S$ is called the support of $\bm{\beta}$, in
other words, $supp(\bm{\beta})=\{i:\bm{\beta}_{i}\neq0\}$. Similarly,
if $\bm{X}$ is a matrix instead of a vector, then $\bm{X}_{S}$ are
the columns of $\bm{X}$ indexed by $S$. Denote the ones vector $\vone_{n}$ as a
$n\times1$ vector with all elements being 1, $\bm{J_{n}}=\vone_{n}\vone_{n}'$,
and $\bm{\bar{J}_{n}}=\frac{1}{n}\bm{J_{n}}$. $\bm{I_{n}}$ denotes
the identity matrix with diagonal 1 and 0 elsewhere. The subscript
$n$ is always omitted when the dimension $n$ is clear from the context.
The notation $\#S$ means the number of elements in the set $S$.

For any matrix $M=\{m_{i,j}\}_{i,j=1}^{n}$ define the following terms
\begin{align}
\text{k-th skew-diagonal} & =\{m_{i,j}|i=j+k\}\label{eq: skew_diag}\\
\text{k-th skew-anti-diagonal} & =\{m_{i,j}|i=n+k+1-j\}\label{eq: skew_anti_diag}
\end{align}
where $-n+1\leq k\leq n-1$.

Because of this high-dimension problem and the high-correlation among
the basis assets, traditional methods fail to give an interpretable
and systematic way to fit the AMF model. Therefore,
we employ the GIBS algorithm proposed in the paper Zhu et al. (2020) \cite{zhu2020high} to select
the basis assets set for each stock .

We give a brief review of the GIBS algorithm in this section. In the
GIBS algorithm, a procedure using the Minimax-Linkage Prototype Clustering
is employed to obtain low-correlated ETFs (denoted as $U$ in the
paper). The high-dimensional statistical methods (the Minimax-Linkage
Prototype Clustering and LASSO) used in the GIBS algorithm can be
found in Appendix \ref{ap: ti_high_dim_stats}. The sketch of the
GIBS algorithm is shown in the Table \ref{tab: ti_gibs_algo}. The
details of the GIBS algorithm can be found in the paper Zhu et al.
(2020) \cite{zhu2020high}. An application can be found in Jarrow et al. (2021) \cite{jarrow2021low}.

For simplicity, denote $V_{1}(t)=B(t)$ as the money market account,
and $V_{2}(t)$ as the market index. Most of the ETFs, $\Delta\bm{V}_{i}$,
are correlated with $\Delta\bm{V}_{2}$, the market portfolio. Although
this is not true for the other 4 Fama-French factors. Therefore, we
first orthogonalize every other basis asset (excluding $\Delta\bm{V}_{1}$
and $\Delta\bm{V}_{2}$) to $\Delta\bm{V}_{2}$. By orthogonalizing
with respect to the market return, we avoid choosing redundant basis
assets and increase the accuracy of the fitting. Note that for Ordinary
Least-Squares (OLS) regression, projection does not affect the estimation
of $\Delta\bm{\hat{Y}}$, since it only affects the coefficients.
However, in LASSO projection does affect the set of selected basis
assets because it changes the magnitude of the coefficients before
shrinking. Thus, we compute
\begin{equation}
\widetilde{\Delta\bm{V}}_{i}=(\bm{I}-P_{\Delta\bm{V}_{2}})\Delta\bm{V}_{i}=(\bm{I}-\Delta\bm{V}_{2}(\Delta\bm{V}'_{2}\Delta\bm{V}_{2})^{-1}\Delta\bm{V}_{2}')\Delta\bm{V}_{i}\qquad\text{where}\quad3\leq i\leq p_{2}\label{eq: ti_proj1}
\end{equation}
where $P$ denotes the projection operator. Let
\begin{equation}
\widetilde{\Delta\bm{V}}=(\bm{V}_{1},\bm{V}_{2},\widetilde{\Delta\bm{V}_{3}},
...,\widetilde{\Delta\bm{V}_{p_{2}}}).
\label{eq: ti_proj2}
\end{equation}
Note that this is equivalent to the residuals after regressing other
risk factors on $\Delta\bm{V}_{2}$.

The transformed ETF basis assets $\widetilde{\Delta\bm{V}}$ still
contain highly correlated members. We first divide these basis assets
into categories $C_{1},C_{2},...,C_{k}$ based on a financial characterization.
Note that $C\equiv\cup_{i=1}^{k}C_{i}=\{1,2,...,p_{1}\}$. The list
of categories with descriptions can be found in Appendix \ref{ap: etf_classes}.
The categories are (1) bond/fixed income, (2) commodity, (3) currency,
(4) diversified portfolio, (5) equity, (6) alternative ETFs, (7) inverse,
(8) leveraged, (9) real estate, and (10) volatility.

Next, from each category we choose a set of representatives. These
representatives should span the categories they are from, but also
have low correlation with each other. This can be done by using the
prototype-clustering method with a distance measure defined in Appendix
\ref{ap: ti_high_dim_stats}, which yield the ``prototypes'' (representatives)
within each cluster (intuitively, the prototype is at the center of
each cluster) with low-correlations.

Within each category, we use the prototype clustering methods previously
discussed to find the set of representatives. The number of representatives
in each category can be chosen according to a correlation threshold.
This gives the sets $D_{1},D_{2},...,D_{k}$ with $D_{i}\subset C_{i}$
for $1\leq i\leq k$. Denote $D\equiv\cup_{i=1}^{k}D_{i}$. Although
this reduction procedure guarantees low-correlation between the elements
in each $D_{i}$, it does not guarantee low-correlation across the
elements in the union $D$. So, an additional step is needed, in which
is prototype clustering on $D$ is used to find a low-correlated representatives
set $U$. Note that $U\subseteq D$. Denote $p_{2}\equiv\#U$.

Recall that $\widetilde{\Delta\bm{V}_{U}}$ means the columns of the
matrix $\widetilde{\Delta\bm{V}}$ indexed by the set $U$. Since
basis assets in $\widetilde{\Delta\bm{V}_{U}}$ are not highly correlated,
a LASSO regression can be applied. Therefore, we have that
\begin{equation}
\bm{\widetilde{\beta}}_{i}=\underset{\bm{\beta}_{i}\in\mathbb{R}^{p},(\bm{\beta}_{i})_{j}=0(\forall j\in U^{c})}{\arg\min}\left\{ \frac{1}{2n}\norm{\Delta\bm{Y}_{i}-\widetilde{\Delta\bm{V}_{U}}\bm{\beta}_{i}}_{2}^{2}+\lambda\norm{\bm{\beta}_{i}}_{1}\right\}
\end{equation}
where $U^{c}$ denotes the complement of $U$. However, here we use
a different $\lambda$ as compared to the traditional LASSO. Normally
the $\lambda$ of LASSO is selected by cross-validation. However this
will overfit the data as discussed in the paper Zhu et al. (2020)
\cite{zhu2020high}. So here we use a modified version of the $\lambda$
selection rule and set
\begin{equation}
\lambda=\max\{\lambda_{1se},min\{\lambda:\#supp(\bm{\widetilde{\beta}}_{i})\leq20\}\}
\end{equation}
where $\lambda_{1se}$ is the $\lambda$ selected by the ``1se rule''.
The ``1se rule'' provides the most regularized model such that the
error is within one standard error of the minimum error achieved by
cross-validation (see \cite{friedman2010regularization,simon2011Regularization,tibshirani2012strong}).
Therefore, the set of basis assets selected is
\begin{equation}
S_{i}\equiv supp(\bm{\widetilde{\beta}}_{i}).
\label{eq: ti_factor_select}
\end{equation}

Next, we fit an Ordinary Least-Square (OLS) regression on the selected
basis assets, to estimate $\bm{\hat{\beta}}_{i}$, the OLS estimator
from
\begin{equation}
\Delta\bm{Y}_{i}=\Delta\bm{V}_{S_{i}}(\bm{\beta}_{i})_{S_{i}}+\Delta\bm{\epsilon}_{i}.\label{eq: ti_ols_subset}
\end{equation}
Since this is an OLS regression, we use the original basis assets
$\Delta\bm{V}_{S_{i}}$ rather than the orthogonalized basis assets
$\widetilde{\Delta\bm{V}}_{S_{i}}$. Note that $supp(\bm{\hat{\beta}}_{i})\subset S_{i}$.
Since we are in the OLS regime, significance tests can be performed
on $\bm{\beta}_{i}$. This yields the significant set of coefficients
\begin{equation}
S_{i}^{*}\equiv\{j:P_{H_{0}}(|\bm{\beta}_{i,j}|\geq|\bm{\hat{\beta}}_{i,j}|)<0.05\}\quad\text{where}\quad H_{0}:\text{True value}\,\,\bm{\beta}_{i,j}=0.
\end{equation}
Note that the significant basis asset set is a subset of the selected
basis asset set. In another words,
\begin{equation}
S_{i}^{*}\subseteq supp(\bm{\hat{\beta}}_{i})\subseteq S_{i}\subseteq\{1,2,...,p\}.
\end{equation}

A sketch of the GIBS algorithm is shown in Table \ref{tab: ti_gibs_algo}.
Recall that for an index set $S\subseteq\{1,2,...,p\}$, $\Delta\bm{V}_{S_{i}}$
means the columns of the matrix $\Delta\bm{V}$ indexed by the set
$S_{i}$.
\begin{table}[H]
\centering
  \begin{tabular}{|| l ||}
  \hhline{|=|}
  \\[-1.6ex]
  \textbf{The Groupwise Interpretable Basis Selection (GIBS) algorithm }  \\[1ex]
  \hhline{|=|}
  \\[-1.6ex]
  Inputs: Stocks to fit $ \Delta\bm{Y}_{i} $ and basis assets $ \Delta\bm{V} $.
  \\[0.6ex]
  \hline
  \\[-1.6ex]
  1. Derive $ \widetilde{\Delta\bm{V}} $ using $ \Delta\bm{V} $ and the Equation (\ref{eq: ti_proj1}, \ref{eq: ti_proj2}).
  \\[0.6ex]
  2. Divide the transformed basis assets $ \widetilde{\Delta\bm{V}} $ into k groups $C_{1}, C_{2},\cdots C_{k}$ using a
  \\[0.6ex]
  \hspace{0.4cm} financial interpretation.
  \\[0.6ex]
  3. Within each group, use prototype clustering to find prototypes $ D_i \subset C_i $.
  \\[0.6ex]
  4. Let $ D = \cup_{i = 1}^k D_i $, use prototype clustering in D to find prototypes $ U \subset D $.
  \\[0.6ex]
  5. For each stock $ \Delta\bm{Y}_{i} $, use a modified version of LASSO to reduce $\bm{\widetilde{X}}_{U}$
  \\[0.6ex]
  \hspace{0.4cm} to the selected basis assets $\widetilde{\Delta\bm{V}_{S_i}}$.
  \\[0.6ex]
  6. For each stock $ \Delta\bm{Y}_{i} $, fit linear regression on $ \Delta\bm{V}_{S_i} $.
  \\[0.6ex]
  \hline
  \\[-1.6ex]
  Outputs: Selected factors $ S_i $, significant factors $ S_i^*$, and coefficients in step 6.
  \\[0.6ex]
  \hhline{|=|}
  \end{tabular}
  \caption{The sketch of Groupwise Interpretable Basis Selection (GIBS) algorithm}
  \label{tab: ti_gibs_algo}
\end{table}

\section{Testing Methodologies and Results}

\label{sec: ti_test_methodology}

This section gives the testing methodologies and results. We first
do an intercept (arbitrage) test, which validates the AMF model we
use. Then we use the indicator variable to test the time-invariance
of the $\beta$'s in a linear setting. After, we do a residual test
to check whether including more basis assets provides a better fit.
At last we compare the fitting with a Generalized Additive Model (GAM)
to test the time-invariance of the $\beta$'s in a non-linear setting.
For each test we repeat it on all the time periods discussed in Section
\ref{sec: ti_data} and report the results.

\subsection{The Intercept Test}

We test the validity of the generalized APT by adding an intercept
(Jensen's alpha) and testing if the intercept is non-zero. A non-zero
intercept implies that the securities are mispriced (i.e. the rejection
of the existence of an equivalent martingale measure). Formally, we
add an intercept term $\alpha_{i}$ to equation (\ref{eq: ti_ols_subset})
and test the null hypothesis
\begin{equation}
H_{0}:\alpha_{i}=0\quad vs.\quad H_{a}:\alpha_{i}\neq0.
\end{equation}
Since using price differences removes the intercept, the intercept
test has to done using prices
\begin{equation}
\bm{Y}_{i}=\alpha_{i}\vone_n+\bm{V}_{S_{i}}(\bm{\beta}_{i})_{S_{i}}+\bm{\epsilon}_{i}.
\label{eq: ti_int}
\end{equation}
where $\vone$ is an $n\times1$ vector with all elements 1, and $S_{i}$
are the basis assets selected by the GIBS algorithm in the AMF model,
defined in Equation (\ref{eq: ti_factor_select}). For the FF5, the
$S_{i}$ are the Fama-French 5-factors and the risk free rate.

Our initial idea was to fit an OLS regression
on the selected basis assets for each company and then report the
p-values for the significance of $\alpha_{i}$. However, we observed
that the mma's value $\bm{V}_{1}$ is highly correlated to the constant
vector $\vone$ because the risk-free rate is close to (or equal to)
0 for a long time. Therefore, including both $\vone$ and $\bm{V}_{1}$
in the regression leads to the inverse of a nearly-singular matrix,
which gives unreliable results. In this case, since the correlation
is so large, even projecting out $\vone$ from $\bm{V}_{1}$ does
not solve this problem. So we used a two-step procedure instead. First,
we estimate the OLS coefficient $(\bm{\hat{\beta}}_{i})_{S_{i}}$
from the the non-intercept model
\begin{equation}
\bm{Y}_{i}=\bm{V}_{S_{i}}(\bm{\beta}_{i})_{S_{i}}+\bm{\epsilon}_{i}
\end{equation}
and calculate the estimation of residuals
\begin{equation}
\bm{\hat{\epsilon}}_{i}=\bm{Y}_{i}-\bm{\hat{Y}}_{i}
\quad\text{where}\quad\bm{\hat{Y}}_{i}=\bm{V}_{S_{i}}(\bm{\hat{\beta}}_{i})_{S_{i}}.
\end{equation}
Then, we fit an intercept-only regression on the residuals
\begin{equation}
\bm{\hat{\epsilon}}_{i}=\alpha_{i}\vone+\bm{\delta}_{i}
\end{equation}
and report the p-value for the significance of the intercept. Using
this technique, we avoided the collinearity issue. The results show
that for all time periods, for all stocks, we can not reject the null
hypothesis in either AMF or FF5. In other words, there is no significant
non-zero intercept for either AMF or FF5 for any time period and for
any stock. This evidence provides a validation of the generalized
APT and the use of the AMF and FF5 models.

\subsection{Time-invariance Test in Linear Setting}

\label{sec: ti_test_linear}

For each time period, this section tests for the time-invariance of
the multi-factors beta coefficients in a linear setting. For these
tests we use the first order differences of the prices as described
in equations (\ref{eq: ti_vec_diff}) and (\ref{eq: ti_ols_subset}).
We use price differences to avoid autocorrelation and any non-stationarities
in the price process. Here we only focus on the selected basis assets.
In other words, we only test the time-invariance of $\beta_{i,j}$
where $j\in S_{i}$ and $S_{i}$ is the one defined in equation (\ref{eq: ti_factor_select}).
Our null hypothesis for each stock $i$ is that ``$H_{0}$: $\beta_{i,j}$
are time-invariant over the 4 years.''

Denote $\bm{h}=(h_{1},h_{2},...,h_{n})'$ where $h_{i}=0$ for all
rows related to first half of the time period and $h_{i}=1$ for all
rows related to the second half of the time period. Testing for the
significance of interaction of each basis asset with $\bm{h}$ is
a way to test whether the coefficients are the same for the first
and last two years. To be more specific, consider the regression model:
\begin{equation}
\Delta\bm{Y}_{i}=\Delta\bm{V}_{S_{i}}(\bm{\beta}_{i})_{S_{i}}+\left[\Delta\bm{V}_{S_{i}}\odot(\bm{h}\vone_{n}')\right]\bm{\theta}_{i}+\Delta\bm{\epsilon}_{i}\label{eq: ti_test_linear}
\end{equation}
Note that the sign ``$\odot$'' means the element-wise multiplication
for two vectors. Here $\theta_{i,j}=0$ indicates that $\beta_{i,j}$
is time-invariant during the time period. Our null hypothesis becomes
\begin{equation}
H_{0}:\theta_{i,j}=0\;\;\text{for}\;\;\forall j\quad vs.\quad H_{1}:\exists j,\;\;\text{such that}\;\;\theta_{i,j}\neq0.
\end{equation}

\begin{figure}[H]
\centering \includegraphics[width=0.9\textwidth]{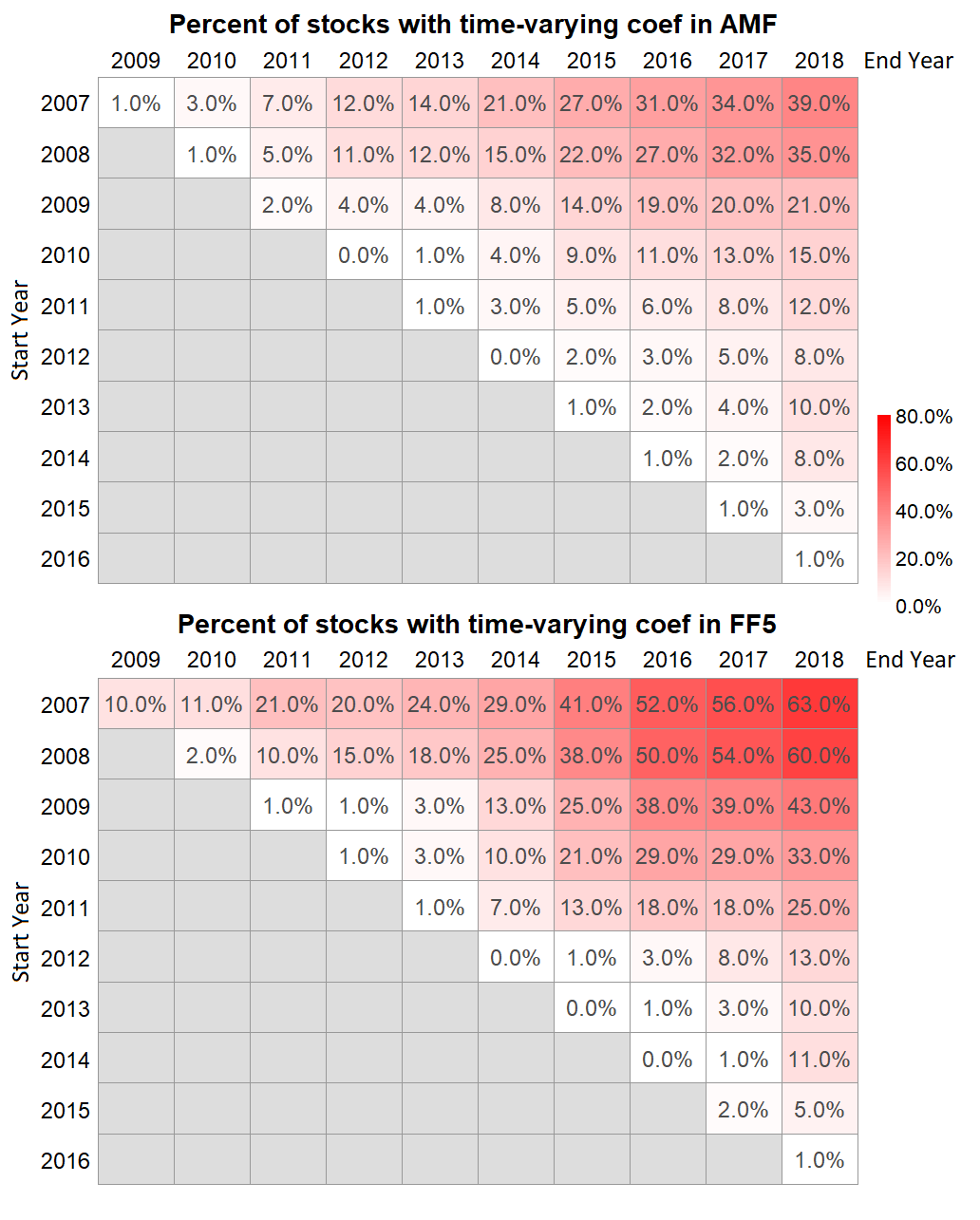} \\
 \vspace{-0.2cm}
 \caption{Percentage of stocks with time-varying beta using the time-invariance
test in a linear setting for each time period. The y-axis is the start
year of each time period and the x-axis is the end year. The percentage
in each grid is the percentage of stocks with FDR Q-threshold 0.05
in Section \ref{sec: ti_test_linear} ANOVA test comparing the models
in Equation (\ref{eq: ti_test_linear}) and (\ref{eq: ti_ols_subset}).}
\label{fig: ti_test_linear_combine}
\end{figure}

An ANOVA test is employed to compare the model in equation (\ref{eq: ti_test_linear})
and (\ref{eq: ti_ols_subset}). A p-value of less than 0.05 rejects
the null hypothesis that the $\beta_{i,j}$'s are all time-invariant.
We also want to control for the False Discovery Rate (FDR) (see \cite{benjamini1995controlling}).
The Benjamini-Hochberg (BH) \cite{benjamini1995controlling} FDR adjusting
procedure does not account for the correlation between tests, while
the Benjamini-Hochberg-Yekutieli (BHY) \cite{benjamini2001control}
FDR adjusting method does. Since we may have correlation between the
basis assets, we use the BHY method to adjust the p-values into the
FDR Q-values and then report the percentage of stocks with Q-values
less than 0.05 in Figure \ref{fig: ti_test_linear_combine}.

Figure \ref{fig: ti_test_linear_combine} reports the percentage of
stocks with time-varying beta using the time-invariance test in a
linear setting for each time period. The y-axis is the start year
of each time period and the x-axis is the end year of the time period.
The percentage in each grid is the percentage of stocks with FDR Q-threshold
0.05 in the ANOVA test comparing the models in Equation (\ref{eq: ti_test_linear})
and (\ref{eq: ti_ols_subset}). The larger the percentage is, the
darker the grid will be. The upper heatmap is the result for the AMF
model, while the bottom heatmap is the result for the FF5 model.

In the heatmaps in Figure \ref{fig: ti_test_linear_combine} (and
same below), all elements on the k-th skew-diagonal (see definition
in Equation (\ref{eq: skew_diag})) correspond to time periods of
the same length, which is $k+3$ years. For example, the diagonal
(0-th skew-diagonal) elements are related to the time periods with
3 years, the 1st skew-diagonal elements are of the time periods 4
years. Comparing the different skew-diagonals, we can see that the
AMF model is very stable in all time periods less than 5 years. For
most time periods no more than 5 years, only less than 5\% companies
has a least 1 time-varying $\beta$. In other words, for more than
95\% of the companies, the $\beta$'s in the AMF model are time-invariant.
However, for FF5 model, even some 3-year time periods are not stable,
such as 2007 - 2009.

In the heatmaps in Figure \ref{fig: ti_test_linear_combine} (and
same below), all elements on the k-th skew-anti-diagonal (see definition
in Equation (\ref{eq: skew_anti_diag})) correspond to time periods
with the same ``mid-year''. For example, the 1st skew-anti-diagonal
elements are all related to time periods centered in the mid week
of 2012, with different time lengths. By comparing different skew-anti-diagonals,
we can compare if the stability pattern are the same for different
mid-years. For AMF model, we can see that the percentage for a fixed
time length does not change much for different mid-years. For example,
all time periods of 4 years has time-varying percent less then 5\%,
which does not change much for different mid-years. However, the stability
of the $\beta$'s for FF5 model depends highly on the mid-year, not
just the length of the time period. FF5 is more volatile in the mid-years
2012 - 2013 and 2008 - 2009. FF5 can not capture the basis assets
as accurately as the AMF, and the $\beta$'s change significantly
during the financial crisis.

\begin{figure}[H]
\centering \includegraphics[width=0.9\textwidth]{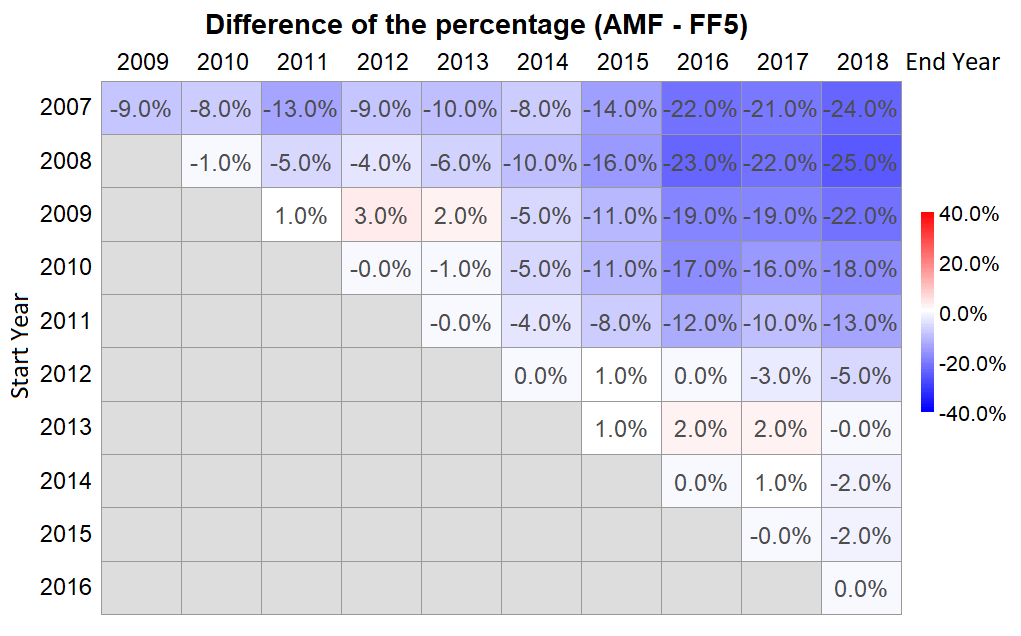} \\
 \vspace{-0.2cm}
 \caption{Difference of the two heatmaps in Figure \ref{fig: ti_test_linear_combine}
to compare AMF and FF5. Each grid is the percent of time-varying stocks
in AMF model minus the percent of time-varying stocks in FF5 model
shown in Figure \ref{fig: ti_test_linear_combine}.}
\label{fig: ti_test_linear_diff}
\end{figure}

In general, the AMF is more stable than the FF5, which can be seen
in Figure \ref{fig: ti_test_linear_diff}. The table in Figure \ref{fig: ti_test_linear_diff}
is the difference between the two tables in Figure \ref{fig: ti_test_linear_combine}
(AMF - FF5). For most time periods, the grid is blue, meaning that
the AMF model is more stable than FF5 by having less a percentage
of companies with time-varying $\beta$'s; sometimes the decrease
is over 20\%. FF5 is slightly more stable than AMF in only a few time
periods, although both AMF and FF5 are quite stable, giving less than
5\% of the companies with time-varying $\beta$'s. AMF performs much
better than FF5 in all other time periods when the FF5 is unstable.

In summary, AMF outperforms the FF5 in terms of the stability in two
ways. First, for all time periods, AMF is more stable (or at least
equally stable) than FF5, i.e. AMF either gives more stable $\beta$'s
than FF5, or gives equally stable $\beta$'s if FF5 is also stable.
Second, the stability of the AMF model for each time length is more
robust across all mid-years as compared to FF5. The stability of the
AMF model only depends on the length of the time period, and not the
starting or ending year, while the stability of the FF5 model depends
on the mid-year. This implies that the AMF performed well during the
financial crisis.

\subsection{Residual analysis}

\label{sec: ti_resid_test}

In this section we test if including more basis assets improves the
fit in each time period. Since the number of ETFs increased over time,
by focusing on the second half of the time period, we have more ETFs
available compared to the beginning. We want to test if the newly
introduced ETFs in the second half of the time period provides a better
fit for the AMF model.

\begin{figure}[H]
\centering \includegraphics[width=0.9\textwidth]{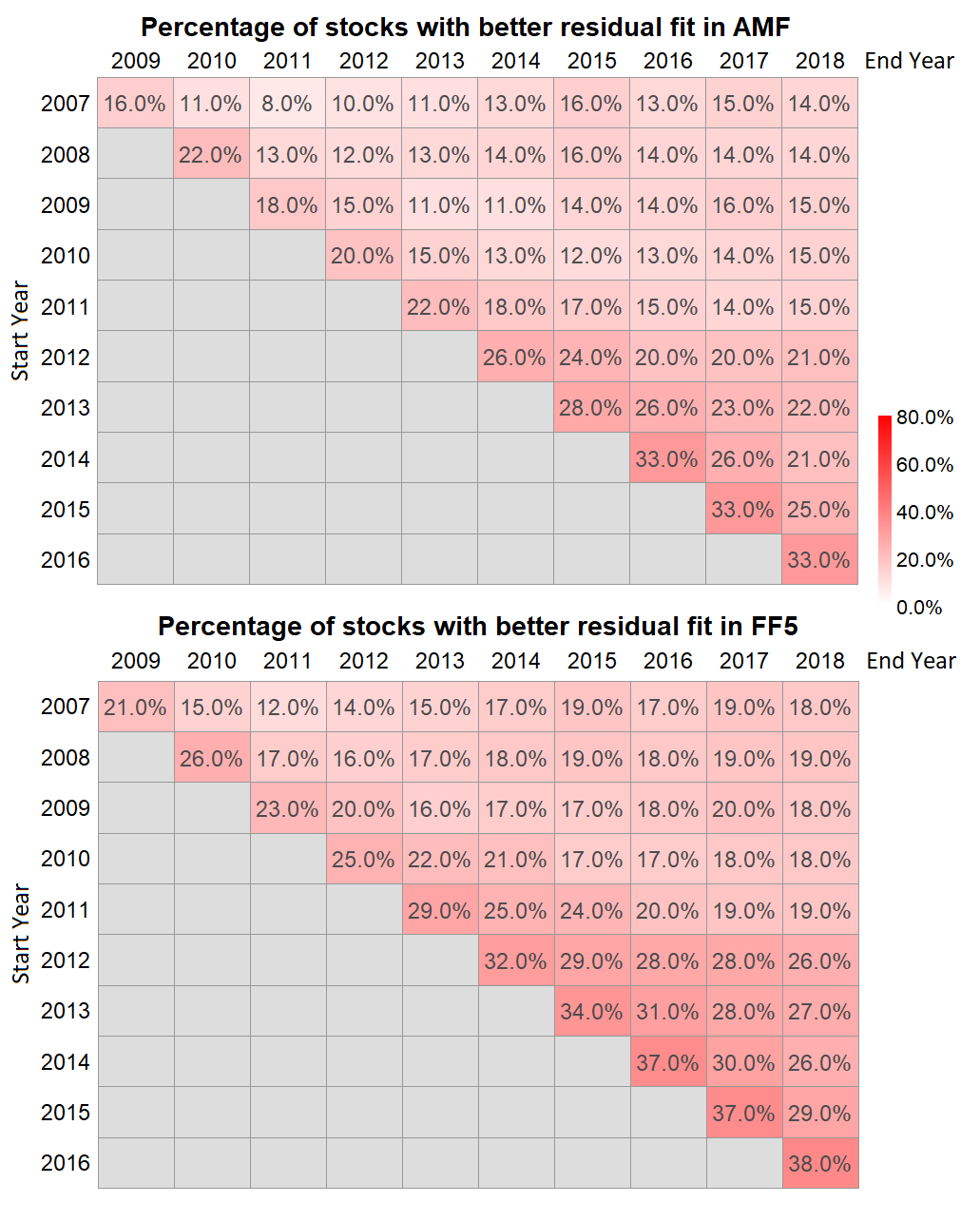}
\\
 \vspace{-0.2cm}
 \caption{Percentage of stocks where the model in Equation (\ref{eq: ti_resid2})
has significantly better fit compared to Equation (\ref{eq: ti_resid2})
for each time period. The y-axis is the start year of each time period
and the x-axis is the end year. The percentage in each grid is the
percentage of stocks with FDR Q-threshold 0.05.}
\label{fig: ti_test_resid_combine}
\end{figure}

Formally, for a time period $[t_{a},t_{b}]$, let $t_{mid}=(t_{a}+t_{b})/2$.
We divide $\Delta\bm{Y}_{i}$ and $\Delta\bm{V}$ into two parts,
one for the time period $[t_{a},t_{mid})$ and the second for the
time period $[t_{mid},t_{b})$, where
\begin{equation}
\Delta\bm{Y}_{i}=\begin{pmatrix}\Delta\bm{Y}_{i,a}\\
\Delta\bm{Y}_{i,b}
\end{pmatrix},\quad\Delta\bm{V}=\begin{pmatrix}\Delta\bm{V}_{a}\\
\Delta\bm{V}_{b}
\end{pmatrix}.
\end{equation}
We first derive the basis assets set $S_{i}$ in Equation (\ref{eq: ti_factor_select})
using the GIBS algorithm on the whole time period. Then, we fit an OLS regression on the second half
\begin{equation}
\Delta\bm{Y}_{i,b}=(\Delta\bm{V}_{b})_{S_{i}}(\bm{\gamma}_{i})_{S_{i}}+\Delta\bm{\epsilon}_{i,b}
\label{eq: ti_resid1}
\end{equation}
and obtain the estimated coefficients $(\bm{\hat{\gamma}}_{i})_{S_{i}}$.
The residuals are
\begin{equation}
\Delta\bm{\hat{\epsilon}}_{i,b}=\Delta\bm{Y}_{i,b}
-(\Delta\bm{V}_{b})_{S_{i}}(\bm{\hat{\gamma}}_{i})_{S_{i}}.
\end{equation}
We fit an AMF model with the GIBS algorithm again, using the residuals
$\Delta\bm{\hat{\epsilon}}_{i,b}$ as our new dependent variables,
using all the basis assets available for the time period $[t_{mid},t_{b})$,
except for the basis assets already selected in $S_{i}$. This AMF
model on the residuals will provide another selected set of basis
assets $S_{i,b}$. If $S_{i,b}\neq\emptyset$, we merge the two sets
together
\begin{equation}
S_{i,union}=S_{i}\cup S_{i,b}.
\end{equation}
Note that since we remove all the basis assets in $S_{i}$ in our
GIBS fitting of the residuals, $S_{i}\cap S_{i,b}=\emptyset$. Continuing,
we fit another OLS regression on the second half of the data using
the basis assets $S_{i,union}$
\begin{equation}
\Delta\bm{Y}_{i,b}=(\Delta\bm{V}_{b})_{S_{i,union}}(\bm{\gamma}_{i})_{S_{i},union}
+\Delta\bm{\epsilon}_{i,b}.
\label{eq: ti_resid2}
\end{equation}
Finally, we use an ANOVA test to compare the
models in Equation (\ref{eq: ti_resid1}) and Equation (\ref{eq: ti_resid2}).
For each time period, this test is performed on all stocks yielding
a list of p-values. As before, we adjust for FDR using the BHY \cite{benjamini2001control}
method and count the number of companies with FDR Q-values less than
0.05. The percentages are reported in Figure \ref{fig: ti_test_resid_combine}.

Note that if for $i$-th stock the set $S_{i,b}=\emptyset$, there
will not be p-value for this stock since the two models in Equation
(\ref{eq: ti_resid1}) and (\ref{eq: ti_resid2}) are the same. So
this stock will not be counted as a company with a FDR Q-value less
than 0.05. However, when presenting the percentage, we use the total
count of companies available in that time period in the denominator.
This generates more conservative percentages.

For the FF5 model, $S_{i}$ is the FF5 5-factors and the risk free
rate. All the remaining steps in the procedure are the same. The results
for FF5 residuals provide a comparison. From Figure \ref{fig: ti_test_resid_combine}
we see that for all time periods, the second half gives a significantly
better fit with the new basis assets for around 15\% of the stocks
in the AMF model. Comparing this result with Figure \ref{fig: ti_test_linear_combine},
although there are stocks that have constant $\beta$'s, their risks
can be better fitted using the new basis assets. This provides an
interesting insight. More basis assets provides better fitting of
the $\epsilon$ errors, although the affects of the old basis assets
are still the same.

\begin{figure}[H]
\centering \includegraphics[width=0.9\textwidth]{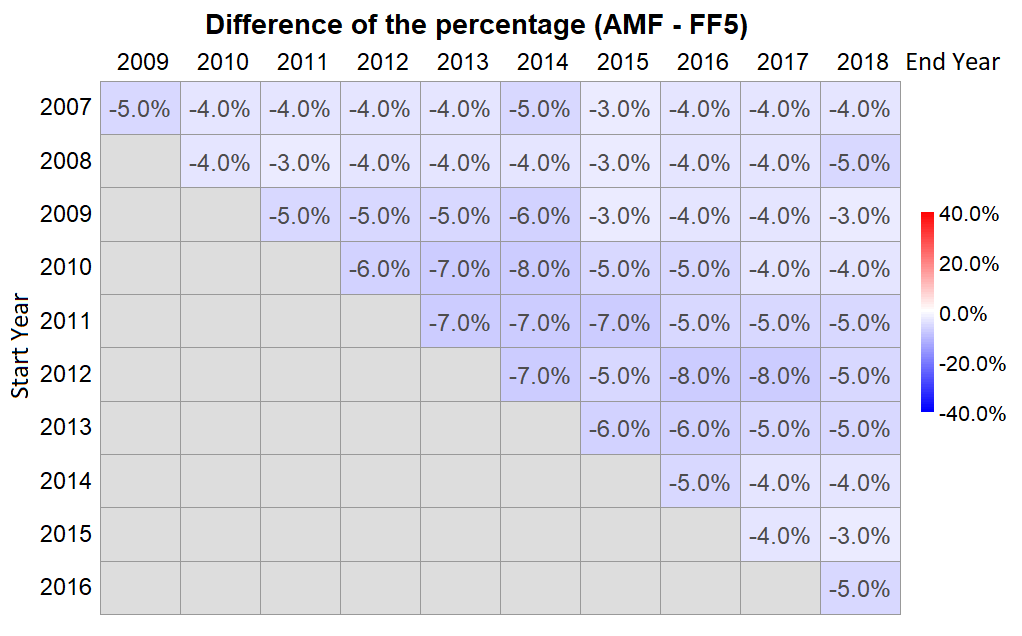} \\
 \vspace{-0.2cm}
 \caption{Difference of the two heatmaps in Figure \ref{fig: ti_test_resid_combine}.
Each grid is the percentage in AMF model minus the percentage in FF5
model shown in Figure \ref{fig: ti_test_resid_combine}.}
\label{fig: ti_test_resid_diff}
\end{figure}

In summary, comparing the results for the AMF and FF5 residuals, it
is clear that AMF uses more basis assets and leaves less information
in the residuals than foes FF5. From Figure \ref{fig: ti_test_resid_diff}
we see that the percentage of stocks that are better fitted in the
second half of the time period is always less for AMF than it is for
FF5.

\subsection{In-Sample and Out-of-Sample Goodness-of-Fit}

\label{sec: gof_test}

This section reports the In-Sample and Out-of-Sample goodness-of-fit
tests. Since the AMF selects more basis assets than the FF5, it is
important to study overfitting. The results show that the AMF model
is more powerful and less vulnerable to overfitting than is the FF5,
because the AMF achieves a better in-sample Adjusted $R^{2}$ (see
\cite{theil1961economic}) and a better Out-of-Sample $R^{2}$ (see
\cite{campbell2008predicting}).

\begin{figure}[H]
\centering \includegraphics[width=0.9\textwidth]{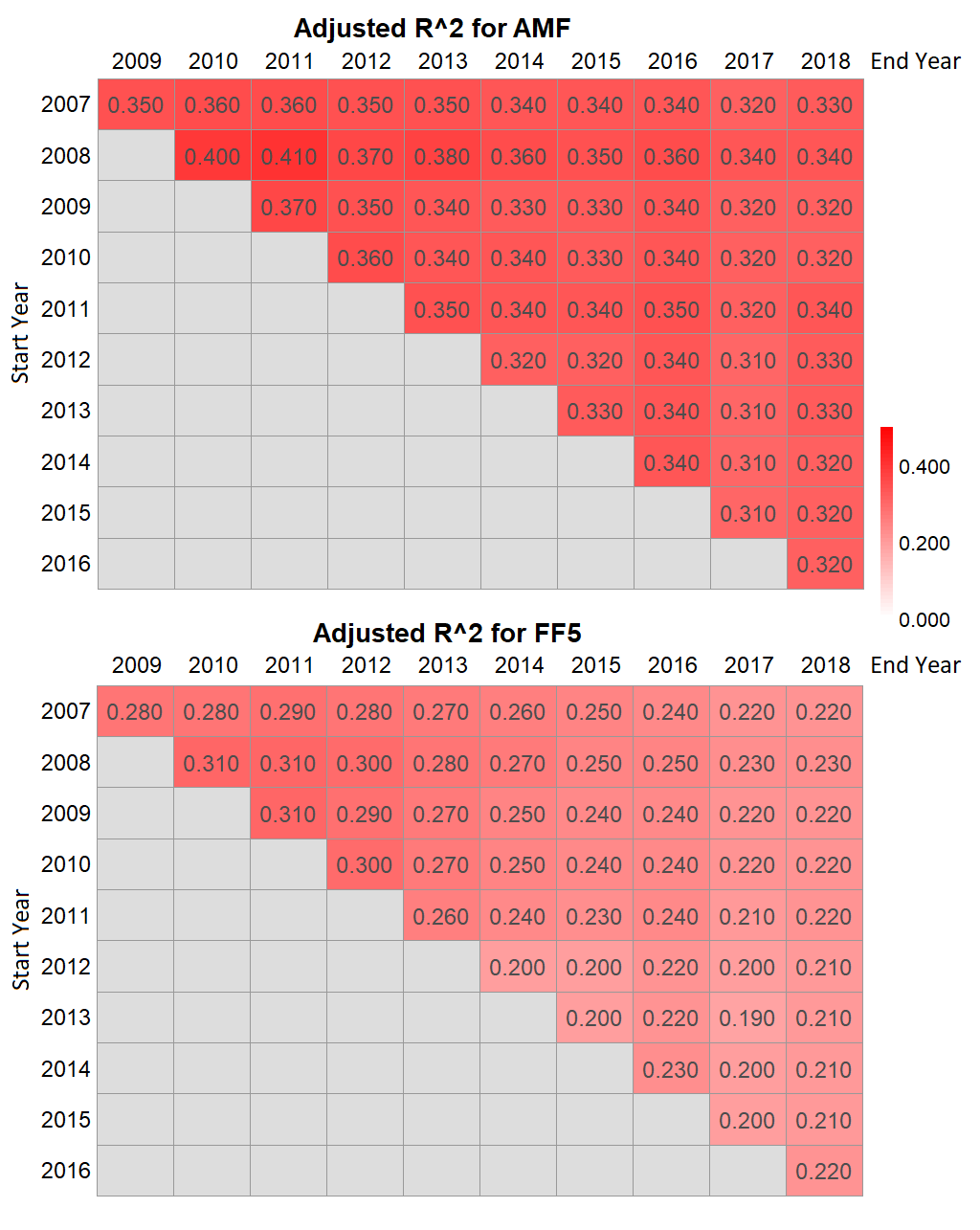} \\
 \vspace{-0.2cm}
 \caption{Mean Adjusted $R^{2}$ for the AMF and FF5 model. The y-axis is the
start year of each time period and the x-axis is the end year. }
\label{fig: ars_combined}
\end{figure}

\begin{figure}[H]
\centering \includegraphics[width=0.9\textwidth]{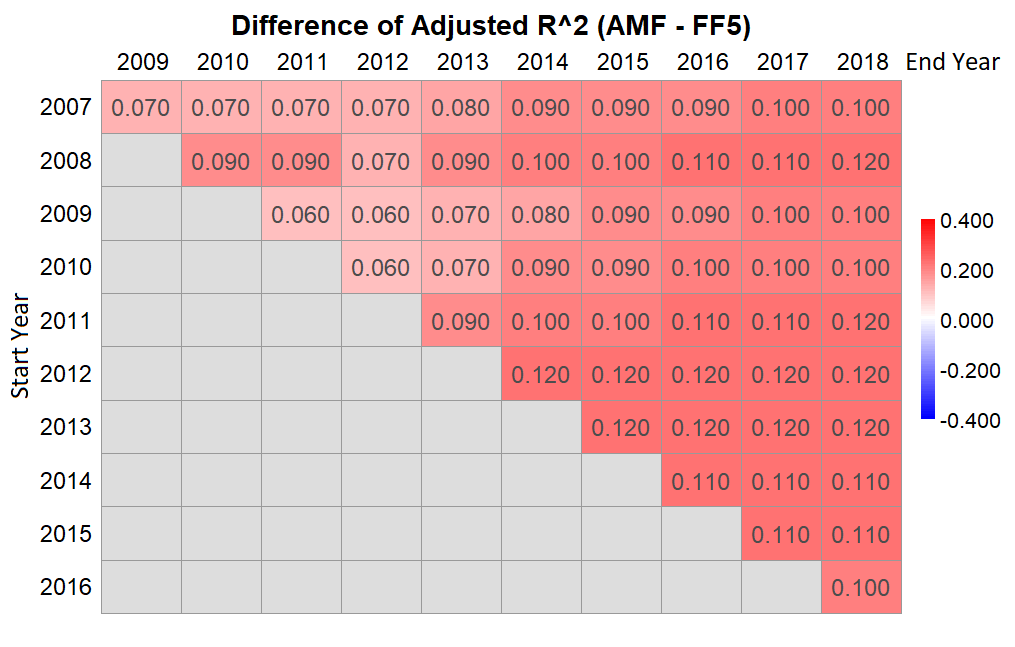} \\
 \vspace{-0.2cm}
 \caption{Difference of the two heatmaps in Figure \ref{fig: ars_combined}.
Each grid is the difference of the mean Adjusted $R^{2}$ of the AMF
model minus that of the FF5 model shown in Figure \ref{fig: ars_combined}.}
\label{fig: ars_diff}
\end{figure}

The Adjusted $R^{2}$ (see \cite{theil1961economic}) is a measure
of goodness of fit similar to the normal $R^{2}$ but adjusted for
the number of parameters to penalize on overfitting. The results are
shown in Figure \ref{fig: ars_combined} and the difference of the
Adjusted $R^{2}$ between AMF and FF5 is shown in Figure \ref{fig: ars_diff}.
It is clear that for all time periods, AMF increases the Adjsuted
$R^{2}$ by around 0.1 compared to FF5, which is around a $0.1/0.25=40\%$
increment.

\begin{figure}[H]
\centering \includegraphics[width=0.9\textwidth]{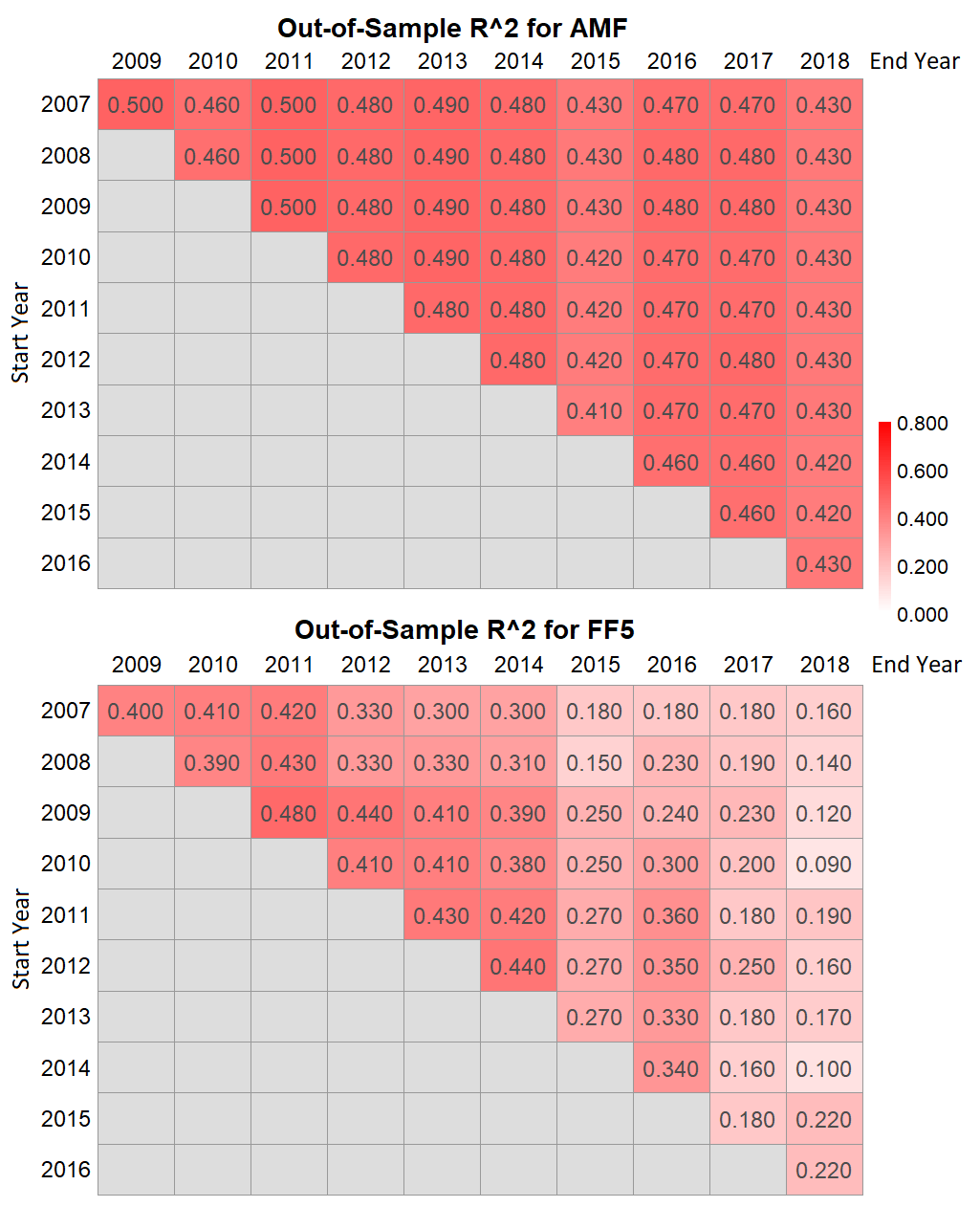}
\\
 \vspace{-0.2cm}
 \caption{Mean Out-of-Sample $R^{2}$ for the AMF and FF5 model. The y-axis
is the start year of each time period and the x-axis is the end year. }
\label{fig: osrs_combined}
\end{figure}

\begin{figure}[H]
\centering \includegraphics[width=0.9\textwidth]{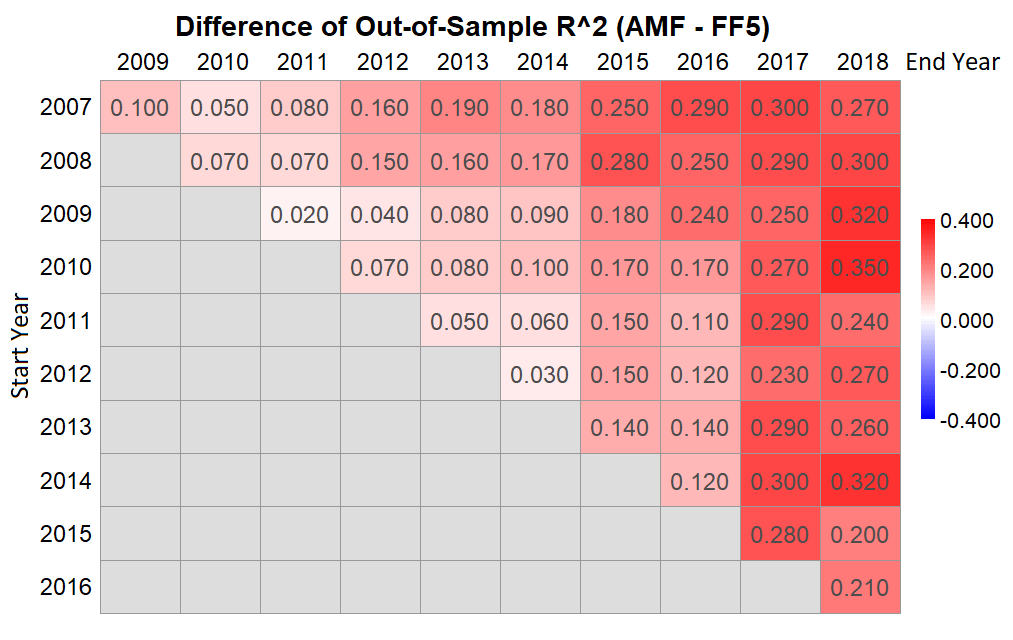} \\
 \vspace{-0.2cm}
 \caption{Difference of the two heatmaps in Figure \ref{fig: osrs_combined}.
Each grid is the difference of the mean Out-of-Sample $R^{2}$ of
the AMF model minus that of the FF5 model shown in Figure \ref{fig: osrs_combined}.}
\label{fig: osrs_diff}
\end{figure}

The Out-of-Sample $R^{2}$ (see \cite{campbell2008predicting}) is
a measure of a model's prediction power. It can be used to check whether
a model is overfitting. For each time period, we calculate the Out-of-Sample
$R^{2}$ for AMF and FF5 for the half-year after the end of that period.
The results are in Figure \ref{fig: osrs_combined} and the differences
of the Out-of-Sample $R^{2}$ between the two models are in Figure
\ref{fig: osrs_diff}. We see that the prediction power of FF5 fades
dramatically with time, while the prediction power of AMF is stable,
and much better than the FF5. The Out-of-Sample $R^{2}$ for AMF is
around 0.47, while that for FF5 is around 0.3 in early years and 0.2
in recent years. In recent years, the prediction power of the AMF
model is almost twice as large as that of FF5. Therefore, it is clear
that the basis assets selected by the AMF model are not overfitting,
but achieving better in-sample and out-of-sample goodness-of-fit.

\subsection{Time-invariance Test in Non-Linear GAM Setting}

\label{sec: ti_test_gam}

We can also test time-invariance based on thr non-linear Generalized
Additive Model (GAM). This is a more strict test, since this tests
not only for the time-invariance of the $\beta$'s, but also their
linearity. We first fit a Time-Varying Coefficient model as a special
case of the GAM. The equation can be
written as:
\begin{equation}
\Delta\bm{Y}_{i}=\Delta\bm{V}_{S_{i}}[\bm{\beta}_{i}(t)]_{S_{i}}+\Delta\bm{\epsilon}_{i}\label{eq: ti_test_gam}
\end{equation}
where $t$ is the time. Note that the only difference between Equation
(\ref{eq: ti_test_gam}) and Equation (\ref{eq: ti_ols_subset}) is
the second allows the $\beta$'s to be functions of time $t$. The
GAM model estimate each $\beta$ as a combination of splines or kernels
with regard to $t$. This can be done by the \textit{gam()} function
in the R package \textit{mgcv}.

\begin{figure}[H]
\centering \includegraphics[width=0.9\textwidth]{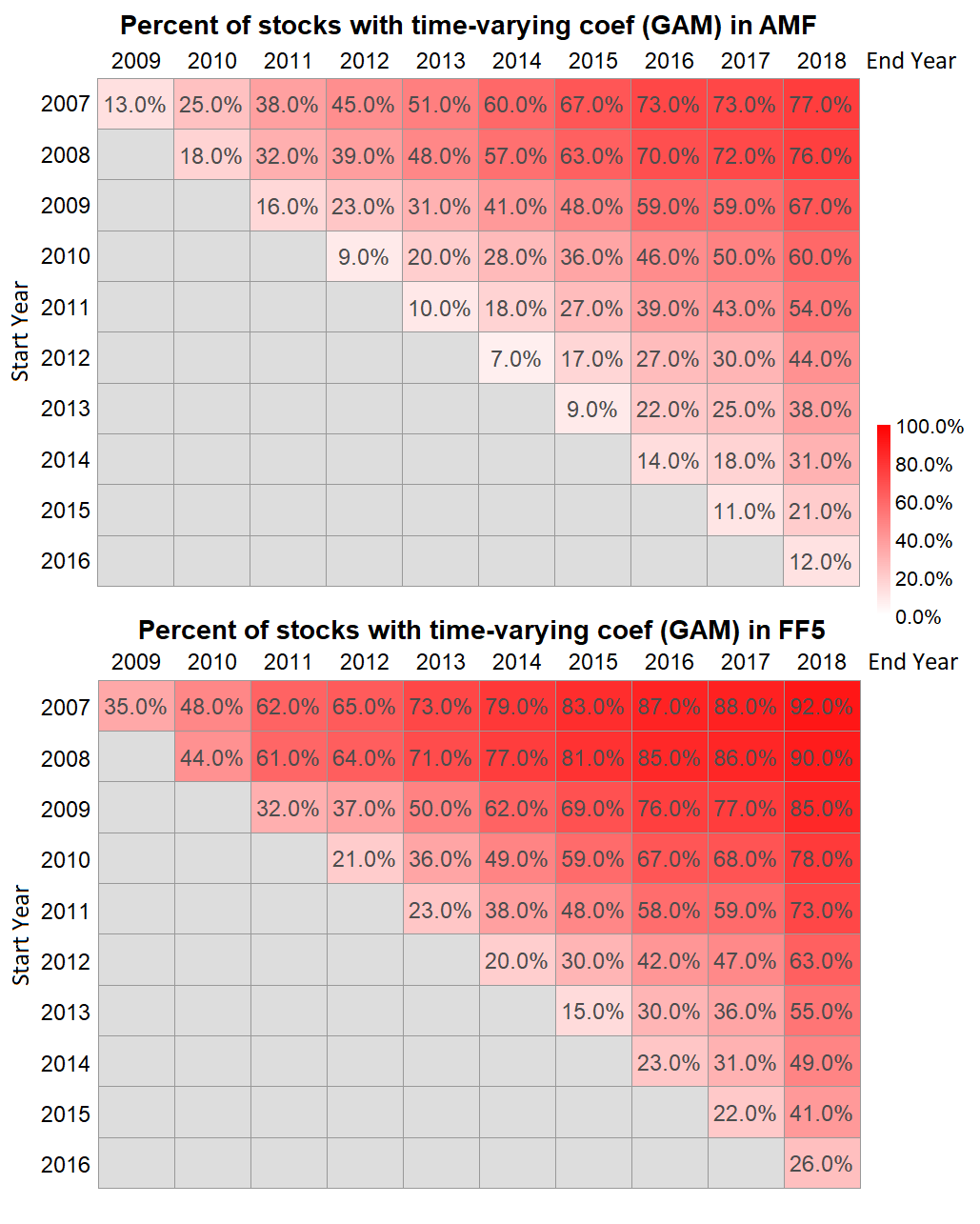} \\
 \vspace{-0.2cm}
 \caption{Percentage of stocks with time-varying beta using the time-invariance
test in a non-linear GAM setting for each time period. The y-axis
is the start year of each time period and the x-axis is the end year.
The percentage in each grid is the percentage of stocks with FDR Q-threshold
0.05 in Section \ref{sec: ti_test_linear} ANOVA test comparing the
models in Equation (\ref{eq: ti_test_gam}) and (\ref{eq: ti_ols_subset}).}
\label{fig: ti_test_gam_combine}
\end{figure}

Next, we perform a ANOVA test between the GAM
model in Equation (\ref{eq: ti_test_gam}) and the linear model in
Equation (\ref{eq: ti_ols_subset}) where each $\bm{\beta}_{i}$ are
constants over time. This yields a p-value for each stock. As before,
we adjust the p-values using the BHY
\cite{benjamini2001control} method to account for FDR. We report the percentage of stocks with Q-values less
than 0.05 in Figure \ref{fig: ti_test_gam_combine}.

Figure \ref{fig: ti_test_gam_combine} reports the percentage of stocks
with time-varying beta using the time-invariance test in a non-linear
GAM setting for each time period. The y-axis is the start year and
the x-axis is the end year of each time period. The percentage in
each grid is the percentage of stocks with FDR Q-threshold 0.05 in
the ANOVA test that compares the models in Equation (\ref{eq: ti_test_gam})
and (\ref{eq: ti_ols_subset}). The larger the percentage, the darker
the grid will be. The upper heatmap is for the AMF model, while the
bottom heatmap is for the FF5 model.

\begin{figure}[H]
\centering \includegraphics[width=0.9\textwidth]{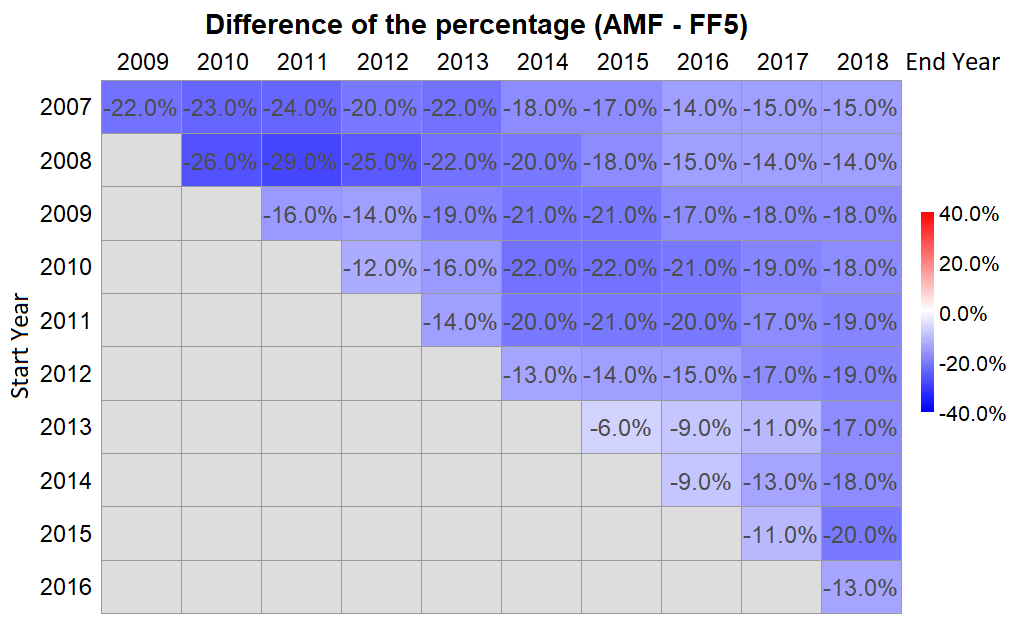} \\
 \vspace{-0.2cm}
 \caption{Difference of the two heatmaps in Figure \ref{fig: ti_test_gam_combine}
to compare AMF and FF5. Each grid is the percent of time-varying stocks
in AMF model minus the percent of time-varying stocks in FF5 model
shown inp Figure \ref{fig: ti_test_gam_combine}.}
\label{fig: ti_test_gam_diff}
\end{figure}

By comparing the different skew-diagonals, we see that both models
are more stable in shorter time periods. The AMF outperforms the FF5
in all the time periods, as seen in Figure \ref{fig: ti_test_gam_diff}.
Figure \ref{fig: ti_test_gam_diff} gives the difference between the
percentages across the two heatmaps in Figure \ref{fig: ti_test_gam_combine}
(AMF - FF5). All the grids are blue, indicating that AMF is more stable
than FF5 in all time periods.

However, comparing the results from GAM and the results from the previous
Section \ref{sec: ti_test_linear}, for any time period and both AMF
and FF5 models, more companies are shown to have time-varying $\beta$'s
in the GAM test, although the AMF still outperforms the FF5. This
does not necessarily mean that the $\beta$'s are time varying in
all time periods, since it is easy for GAM to overfit, especially
in shorter time periods. Indeed, the number of observations for 3
year time periods is $n=156$. For the FF5 model, there are 5 basis
assets. For the AMF model, there are more basis assets selected. For
each basis asset, the GAM model selects some splines or kernels, say
10 splines, which can easily expand the dimension of parameters to
$p=5\times10=50$, which is already too large for a regression with
$n=156$ observations. The number of parameters becomes too close
to the number of observations. In addition, these new variables may
be highly correlated, making the fitt more unstable. This can result
in the GAM overfitting. Therefore, the testing based on the GAM model
in this section is explorative. Considering this high-dimensional
issue, penalization and constraints need to be introduced to traditional
GAM for our application. This extension is left for future research.

\section{Conclusion}

\label{sec: ti_conclude}

The purpose of this paper is to test the multi-factor beta model implied by the Generalized APT and the AMF model with the GIBS algorithm, without imposing the exogenous assumption of constant betas. The intercept (arbitrage) tests show that there is no significant non-zero intercepts in either AMF or FF5 model, which validates the 2 models. The in-sample and out-of-sample goodness-of-fit results show that AMF achieves both better in-sample adjusted $R^{2}$ and out-of-sample $R^{2}$, indicating that AMF is more powerful in fitting and less vulnerable to overfitting.

We perform time-invariance tests for the $\beta$'s for both the AMF
and the FF5 models in various time periods. We show that the constant-beta
assumption holds in the AMF model in all time periods with length
less than 6 years and is quite robust regardless of the start year.
However, even for short time periods, FF5 sometimes gives very unstable
estimation, especially in financial crisis. This indicates that the
AMF is more descriptive and it captures the basis assets that explain
the market movements during the financial crisis. For time periods
with length longer than 6 years, both AMF and FF5 fail to provide
time-invariance $\beta$'s. However, the $\beta$'s estimate by the
AMF is more time-invariant than the FF5 for nearly all time periods.
This shows the superior performance of the AMF model. In summary,
using the dynamic AMF model with a descent rolling window (such as
5 years) is more powerful and stable than is the FF5 model.

%\printbibliography

\bibliographystyle{plain}
\bibliography{bib_liao}

\begin{appendices}

% \clearpage
\section{Prototype Clustering and LASSO}

\label{ap: ti_high_dim_stats}

This section describes the high-dimensional statistical methodologies
used in the GIBS algorithm, including the prototype clustering and
the LASSO. To remove unnecessary independent variables using clustering
methods, we classify them into similar groups and then choose representatives
from each group with small pairwise correlations. First, we define
a distance metric to measure the similarity between points (in our
case, the returns of the independent variables). Here, the distance
metric is related to the correlation of the two points, i.e.
\begin{equation}
d(r_{1},r_{2})=1-|corr(r_{1},r_{2})|
\end{equation}
where $r_{i}=(r_{i,t},r_{i,t+1},...,r_{i,T})'$ is the time series
vector for independent variable $i=1,2$ and $corr(r_{1},r_{2})$
is their correlation. Second, the distance between two clusters needs
to be defined. Once a cluster distance is defined, hierarchical clustering
methods (see \cite{kaufman2009finding}) can be used to organize the
data into trees.

In these trees, each leaf corresponds to one of the original data
points. Agglomerative hierarchical clustering algorithms (e.g. \cite{kaufman2009finding} \cite{huang2020time}) build trees in a
bottom-up approach, initializing each cluster as a single point, then merging
the two closest clusters at each successive stage. This merging is
repeated until only one cluster remains. Traditionally, the distance
between two clusters is defined as either a complete distance, single
distance, average distance, or centroid distance. However, all of
these approaches suffer from interpretation difficulties and inversions
(which means parent nodes can sometimes have a lower distance than
their children), see Bien, Tibshirani (2011)\cite{bien2011hierarchical}.
To avoid these difficulties, Bien, Tibshirani (2011)\cite{bien2011hierarchical}
introduced hierarchical clustering with prototypes via a minimax linkage
measure, defined as follows. For any point $x$ and cluster $C$,
let
\begin{equation}
d_{max}(x,C)=\max_{x'\in C}d(x,x')
\end{equation}
be the distance to the farthest point in $C$ from $x$. Define the
\emph{minimax radius} of the cluster $C$ as
\begin{equation}
r(C)=\min_{x\in C}d_{max}(x,C)
\end{equation}
that is, this measures the distance from the farthest point $x\in C$
which is as close as possible to all the other elements in C. We call
the minimizing point the \emph{prototype} for $C$. Intuitively, it
is the point at the center of this cluster. The \emph{minimax linkage}
between two clusters $G$ and $H$ is then defined as
\begin{equation}
d(G,H)=r(G\cup H).
\end{equation}
Using this approach, we can easily find a good representative for
each cluster, which is the prototype defined above. It is important
to note that minimax linkage trees do not have inversions. Also, in
our application as described below, to guarantee interpretable and
tractability, using a single representative independent variable is
better than using other approaches (for example, principal components
analysis (PCA)) which employ linear combinations of the independent
variables.

The LASSO method was introduced by Tibshirani (1996) \cite{tibshirani1996regression}
for model selection when the number of independent variables ($p$)
is larger than the number of sample observations ($n$). The method
is based on the idea that instead of minimizing the squared loss to
derive the OLS solution for a regression,
we should add to the loss a penalty on the absolute value of the coefficients
to minimize the absolute value of the non-zero coefficients selected.
To illustrate the procedure, suppose that we have a linear model
\begin{equation}
\bm{y}=\bm{X\beta}+\bm{\epsilon}\qquad\text{where}\qquad\bm{\epsilon}\sim N(0,\sigma_{\epsilon}^{2}\bm{I}),
\end{equation}
$\bm{X}$ is an $n\times p$ matrix, $\bm{y}$ and $\bm{\epsilon}$
are $n\times1$ vectors, and $\bm{\beta}$ is a $p\times1$ vector.

The LASSO estimator of $\bm{\beta}$ is given by
\begin{equation}
\bm{\hat{\beta}}_{\lambda}=\underset{\bm{\beta}\in\mathbb{R}^{p}}{\arg\min}\left\{ \frac{1}{2n}\norm{\bm{y}-\bm{X\beta}}_{2}^{2}+\lambda\norm{\bm{\beta}}_{1}\right\}
\end{equation}
where $\lambda>0$ is the tuning parameter, which determines the magnitude
of the penalty on the absolute value of non-zero $\beta$'s. In this
paper, we use the R package \textit{glmnet} \cite{friedman2010regularization}
to fit LASSO.

In the subsequent estimation, we will only use a modified version
of LASSO as a model selection method to find the collection of important
independent variables. After the relevant basis assets are selected,
we use a standard Ordinary Least-Square (OLS) regression on these
variables to test for the goodness of fit and significance of the
coefficients. More discussion of this approach can be found in Zhao,
Shojaie, Witten (2017) \cite{zhao2017defense}.

In this paper, we fit the prototype clustering followed by a LASSO
on the prototype basis assets selected. The theoretical justification
for this approach can be found in Reid et al. (2018) \cite{reid2018general} and Zhao et al. (2017) \cite{zhao2017defense}.

\section{ETF Classes and Subclasses}

\label{ap: etf_classes}

ETFs can be divided into 10 classes, 73 subclasses (categories) in
total, based on their financial explanations. The classify criteria
are found from the ETFdb database: www.etfdb.com. The classes and
subclasses are listed below:
\begin{enumerate}
\item \textbf{Bond/Fixed Income}: California Munis, Corporate Bonds, Emerging
Markets Bonds, Government Bonds, High Yield Bonds, Inflation-Protected
Bonds, International Government Bonds, Money Market, Mortgage Backed
Securities, National Munis, New York Munis, Preferred Stock/Convertible
Bonds, Total Bond Market.
\item \textbf{Commodity}: Agricultural Commodities, Commodities, Metals,
Oil \& Gas, Precious Metals.
\item \textbf{Currency}: Currency.
\item \textbf{Diversified Portfolio}: Diversified Portfolio, Target Retirement
Date.
\item \textbf{Equity}: All Cap Equities, Alternative Energy Equities, Asia
Pacific Equities, Building \& Construction, China Equities, Commodity
Producers Equities, Communications Equities, Consumer Discretionary
Equities, Consumer Staples Equities, Emerging Markets Equities, Energy
Equities, Europe Equities, Financial Equities, Foreign Large Cap Equities,
Foreign Small \& Mid Cap Equities, Global Equities, Health \& Biotech
Equities, Industrials Equities, Japan Equities, Large Cap Blend Equities,
Large Cap Growth Equities, Large Cap Value Equities, Latin America
Equities, MLPs (Master Limited Partnerships), Materials, Mid Cap Blend
Equities, Mid Cap Growth Equities, Mid Cap Value Equities, Small Cap
Blend Equities, Small Cap Growth Equities, Small Cap Value Equities,
Technology Equities, Transportation Equities, Utilities Equities,
Volatility Hedged Equity, Water Equities.
\item \textbf{Alternative ETFs}: Hedge Fund, Long-Short.
\item \textbf{Inverse}: Inverse Bonds, Inverse Commodities, Inverse Equities,
Inverse Volatility.
\item \textbf{Leveraged}: Leveraged Bonds, Leveraged Commodities, \\
 Leveraged Currency, Leveraged Equities, Leveraged Multi-Asset, Leveraged
Real Estate, Leveraged Volati-lity.
\item \textbf{Real Estate}: Global Real Estate, Real Estate.
\item \textbf{Volatility}: Volatility.
\end{enumerate}
\end{appendices}

\end{document}